\documentclass[twocolumn,astrosymb]{aastex701}

\usepackage{lineno}
\usepackage{xspace}
\usepackage{soul}
\usepackage{color}
\usepackage{xcolor}
\usepackage{booktabs}
\usepackage{longtable}

\linenumbers

\shorttitle{ULGRBs and LFBOTs from mergers}
\shortauthors{Villar et al.}

\begin{document}
\title{Compact Objects Merging with Stars as an Origin of Ultra-Long Gamma-Ray Bursts and Luminous Fast Blue Optical Transients}

\correspondingauthor{V.~A.~Villar}
\email{ashleyvillar@cfa.harvard.edu}

\newcommand{\NU}{\affiliation{Center for Interdisciplinary Exploration and Research in Astrophysics (CIERA) and Department of Physics and Astronomy, Northwestern University, Evanston, IL 60208, USA}}

\newcommand{\Purdue}{\affiliation{Purdue University, 
Department of Physics and Astronomy, 525 Northwestern Avenue, West Lafayette, IN 47907, USA}}

\newcommand{\CfA}{\affiliation{Center for Astrophysics\:$|$\:Harvard \& Smithsonian, 60 Garden St. Cambridge, MA 02138, USA}}

\newcommand{\UCSC}{\affiliation{Department of Astronomy and Astrophysics, University of California, Santa Cruz, CA 95064, USA}}

\newcommand{\IS}{\affiliation{Centre for Astrophysics and Cosmology, Science Institute, University of Iceland, Dunhagi 5, 107 Reykjav\'ik, Iceland}}

\newcommand{\DAWN}{\affiliation{Cosmic Dawn Center (DAWN), Niels Bohr Institute, University of Copenhagen, Jagtvej 128, 2100 Copenhagen \O, Denmark}}

\newcommand{\PUCV}{\affiliation{Instituto de F\'isica, Pontificia Universidad Cat\'olica de Valpara\'iso, Casilla 4059, Valpara\'iso, Chile}}

\newcommand{\IPMU}{\affiliation{Kavli Institute for the Physics and Mathematics of the Universe (Kavli IPMU), 5-1-5 Kashiwanoha, Kashiwa, 277-8583, Japan}}

\newcommand{\PSU}{\affiliation{Department of Astronomy \& Astrophysics, The Pennsylvania State University, University Park, PA 16802, USA}}

\newcommand{\ICDS}{\affiliation{Institute for Computational \& Data Sciences, The Pennsylvania State University, University Park, PA, USA}}

\newcommand{\IGC}{\affiliation{Institute for Gravitation and the Cosmos, The Pennsylvania State University, University Park, PA 16802, USA}}

\newcommand{\Swin}{\affiliation{ Centre for Astrophysics and Supercomputing, Swinburne University of Technology, Hawthorn, VIC, 3122, Australia}}

\newcommand{\Curtin}{\affiliation{ International Centre for Radio Astronomy Research, Curtin University, Bentley, WA 6102, Australia}}

\newcommand{\MQ}{\affiliation{Department of Physics \& Astronomy, Macquarie University, NSW 2109, Australia}}

\newcommand{\MQAAAstro}{\affiliation{Macquarie University Research Centre for Astronomy, Astrophysics \& Astrophotonics, Sydney, NSW 2109, Australia}}

\newcommand{\CSIRO}{\affiliation{CSIRO, Space and Astronomy, PO Box 76, Epping NSW 1710 Australia}}

\newcommand{\KICP}{\affiliation{Kavli Institute for Cosmological Physics, The University of Chicago, 5640 South Ellis Avenue, Chicago, IL 60637, USA}}

\newcommand{\UChicago}{\affiliation{Department of Astronomy \& Astrophysics, University of Chicago, 5640 S Ellis Avenue, Chicago, IL 60637, USA}}

\newcommand{\UA}{\affiliation{University of Arizona, Steward Observatory, 933~N.~Cherry~Ave., Tucson, AZ 85721, USA}}

\newcommand{\EFI}{\affiliation{Enrico Fermi Institute, The University of Chicago, 933 East 56th Street, Chicago, IL 60637, USA}}

\newcommand{\mpia}{\affiliation{Max-Planck-Institut f\"{u}r Astronomie (MPIA), K\"{o}nigstuhl 17, 69117 Heidelberg, Germany}}

\newcommand{\GWU}{\affiliation{Department of Physics, The George Washington University, Washington, DC 20052, USA}}

\newcommand{\UCB}{\affiliation{Department of Astronomy, University of California, Berkeley, CA 94720-3411, USA}}

\newcommand{\RU}{\affiliation{Department of Astrophysics/IMAPP, Radboud University, PO Box 9010,
6500 GL, The Netherlands}}

\newcommand{\LJMU}{\affiliation{Astrophysics Research Institute, Liverpool John Moores University, IC2, Liverpool Science Park, 146 Brownlow Hill, Liverpool L3 5RF, UK}}

\newcommand{\LU}{\affiliation{School of Physics and Astronomy, University of Leicester, University Road, Leicester. LE1 7RH, UK}}

\newcommand{\Adler}{\affiliation{The Adler Planetarium, 1300 South DuSable Lake Shore Drive, Chicago, IL 60605, USA}}

\newcommand{\ANU}{\affiliation{Research School of Astronomy and Astrophysics, Australian National University, Canberra, ACT 2611, Australia}}

\newcommand{\Car}{\affiliation{Cardiff Hub for Astrophysics Research and Technology, School of Physics \& Astronomy, Cardiff University, Queen's Buildings, Cardiff CF24 3AA, UK}}

\newcommand{\IAIFI}{\affiliation{The NSF AI Institute for Artificial Intelligence and Fundamental Interactions}}

\newcommand{\MIT}{\affiliation{Department of Physics and Kavli Institute for Astrophysics and Space Research, Massachusetts Institute of Technology, 77 Massachusetts Avenue, Cambridge, MA 02139, USA}}

\newcommand{\Hawaii}{\affiliation{Institute for Astronomy, University of Hawai‘i, 640 N. A‘ohoku Pl., Hilo, HI 96720, USA}}

\newcommand{\Weizmann}{\affiliation{Department of Particle Physics and Astrophysics, Weizmann Institute of Science, 234 Herzl St, 7610001 Rehovot, Israel}}

\newcommand{\Minnesota}{\affiliation{School of Physics and Astronomy, University of Minnesota, Minneapolis, MN 55455, USA}}

\newcommand{\CCA}{\affiliation{Center for Computational Astrophysics, Flatiron Institute, 162 W. 5th Avenue, New York, NY 10011, USA}}

\newcommand{\Columbia}{\affiliation{Department of Physics and Columbia Astrophysics Laboratory, Columbia University, New York, NY 10027, USA}}

\newcommand{\LSU}{\affiliation{Department of Physics \& Astronomy, Louisiana State University, Baton Rouge, LA 70803, USA}}

\newcommand{\LANL}{\affiliation{Center for Nonlinear Studies, Los Alamos National Laboratory, Los Alamos, NM 87545 USA}}

\newcommand{\mpa}{\affiliation{Max Planck Institute for Astrophysics, Karl-Schwarzschild-Strasse 1, 85748 Garching, Germany}}

\author[0000-0002-5814-4061]{V.~Ashley Villar}
\email{ashleyvillar@cfa.harvard.edu}
\CfA
\IAIFI

\author[0000-0002-2028-9329]{Anya E. Nugent}
\email{}
\CfA

\author[0000-0002-2942-3379]{Eric Burns}
\email{}
\LSU

\author[0000-0002-4670-7509]{Brian D.~Metzger}
\email{}
\CCA
\Columbia

\author[0000-0003-0307-9984]{Tarraneh Eftekhari}
\email{}
\NU

\author[0000-0002-7527-5741]{Jakub Klencki}
\email{}
\mpa

\author[0000-0003-2624-0056]{Christopher~L.~Fryer}
\email{}
\LANL
\GWU

\begin{abstract}
Ultra-long gamma-ray bursts (ULGRBs) and luminous fast blue optical transients (LFBOTs) are two rare classes of engine-driven transients whose physical connection remains unknown. It has been suggested that both may arise from the mergers of a massive helium core with a compact object. We investigate this common origin by reanalyzing the optical counterpart of the highly unusual GRB~111209A/SN~2011kl associated with an ULGRB in the context of a recently developed, analytical LFBOT model. We find that SN~2011kl is broadly consistent with an LFBOT origin, exhibiting a rapid, luminous and blue early emission. However, compared to the LFBOT population, SN~2011kl features a longer ``plateau" of emission $\sim2$ weeks post-merger, suggesting an extended pre-merger mass-loss history, as well as stronger UV suppression. We additionally compare the host galaxy environments of five ULGRBs to those of LFBOTs and classical LGRBs. We find that ULGRBs, similar to LFBOTs and long GRBs, tend to occur in lower mass ($<10^{10}$~$M_\odot$) galaxies with higher amounts of active star formation than observed for field galaxy populations at similar redshifts. Together, these results support a shared progenitor for at least a subset of ULGRBs and LFBOTs.
\end{abstract}
\keywords{Gamma-ray bursts (629), Supernovae (1668), Massive stars (732), Close binary stars (254)}

\section{Introduction} \label{sec:intro} 
Ultra-long gamma-ray bursts (ULGRBs) are a rare class of high-energy transients with exceptional prompt gamma-ray durations, $T_{90}\gtrsim$1\, hr, where $T_{90}$ is the central time over which 90\% of the gamma-ray and hard X-ray fluence is observed. The observed duration is thought to broadly follow the timescale of the central engine, with ULGRBs representing systems in which the engine remains active for orders of magnitude longer than classical long GRBs (LGRBs). However, there is a wide diversity of events with $T_{90}\gtrsim10^3$\,s. Some, such as XRF~060218/SN~2006aj \citep{mazzali2006neutron,modjaz2006early} and more recently EP~250108a/SN~2005kg \citep{eyles2025kangaroo,srinivasaragavan2025ep250108a,rastinejad2025ep}, are characterized by low gamma-ray luminosity and soft X-ray spectra. In contrast to low-luminosity GRBs, there exists a distinct population of ULGRBs whose luminosities are more consistent with an extension of the classical LGRB population \citep{virgili2013grb,levan2014new}. In this Letter, we use ULGRB to refer to these high-luminosity events.

Proposed ULGRB progenitors have ranged from the collapse of blue supergiants \citep{gendre2013ultra,tsuna2025fates,chrimes2026luminous}, supramassive magnetars \citep{fu2024grb}, white dwarf tidal disruption events \citep{ioka2016ultra}, and the merger of a helium core/Wolf-Rayet star with a compact object (He-CO; \citealt{fryer1998helium, zhang2001merger, thone2011unusual, fryer2025explaining}). In the He-CO merger scenario, the orbital angular momentum spins up the He core as the binary inspirals, ultimately leading to a CO within the rotating stellar core. The GRB is then powered by accretion onto the CO following the merger, with the duration of the GRB set by a combination of the accretion time of the disk and the freefall time of the stellar material. \cite{fryer2025explaining} find that high orbital angular momentum can substantially extend the accretion timescale compared to that of a single-star system. In particular, ULGRBs with durations $\gtrsim1$ks can be produced by relatively lower black-hole forming zero-age main sequence stars ($M_\mathrm{ZAMS}\sim20M_\odot$) due to long-lived accretion disks. This further suggests that ULGRBs may be less sensitive to metallicity than classical LGRBs and likely lead to notably higher circumstellar densities. The geometry of the circumstellar environment at the time of merger is unclear and may depend on factors such as metallicity, the degree of envelope stripping and the time elapsed since common-envelope ejection (e.g., \citealt{metzger2022luminous}). 

Very few ULGRBs have well characterized, thermal counterparts. GRB~111209A/SN~2011kl is one of the best-studied examples \citep{gendre2013ultra,greiner2015very}. SN~2011kl is more luminous, bluer and faster evolving relative to classical GRB-supernovae (GRB-SNe) such as the archetypical SN~1998bw associated with GRB~980425 \citep{galama1998unusual}. These differences may arise naturally if ULGRBs are powered by longer-lived central engines, with optical light curves reflecting this engine and not dominated by radioactive decay. In general, a roughly fixed fraction of the engine power is channeled into jets versus a thermalized outflow, longer-lived engines may naturally produce brighter optical counterparts because a larger fraction of the energy from the engine can escape via radiation rather than being converted into bulk kinetic energy (and similarly, longer-lived jets should have a lower peak power given a fixed energy reservoir; e.g., \citealt{metzger2015diversity}).

In the optical, time-domain surveys have independently identified a growing population of Luminous Fast Blue Optical Transients (LFBOTs; e.g., \citealt{drout2014rapidly,pursiainen2018rapidly,perley2019fast,coppejans2020mildly}). LFBOTs are characterized by bright peak luminosities (M$\lesssim-20$) with featureless, blue spectral energy distributions that defy standard radioactive decay models of SNe. The progenitors of LFBOTs remain unknown, but X-ray observations suggest the presence of a long-lived, accreting central engine \citep{margutti2019embedded,ho2023minutes,aj2025most,liu20262024wpp}, while radio observations indicate an extended circumstellar medium (CSM; \citealt{margutti2019embedded,ho2020koala,sevilla2026multiwavelength}). A recent analysis of LFBOT host galaxy environments showed that LFBOTs trace lower metallicity galaxies than Type II and ``stripped"  Type Ib/c core-collapse SNe, but higher metallicity environments than LGRBs and superluminous SNe \citep{nugent_lfbots}.

Like ULGRBs, He-CO mergers have been theorized as a progenitor system of LFBOTs \citep{metzger2022luminous,klencki2025luminous}. In this scenario, the optical emission is powered by reprocessing of an inner X-ray source/engine by a fast, polar outflow and interaction between the ejecta and surrounding CSM. Such an outflow naturally explains the featureless spectra observed in LFBOTs sustaining large optical depth despite high expansion velocities \citep{aspegren2026emission,liu20262024wpp}.  If the CO is rapidly spinning at the time of merger, an ULGRB may be launched and observed when on-axis. So while not all He-CO mergers with LFBOTs result in ULGRBs, we should expect that (at least some) ULGRBs have LFBOT-like counterparts. Interestingly, a potential connection between ULGRBs and LFBOTs has also been proposed in an entirely different context, namely the core collapse of rapidly rotating supergiant merger products \citep{tsuna2025fates}; similarities between these populations may not be unique to the He-CO merger scenario.

In this work, we reanalyze the UV through optical light curve of GRB~111209A/SN~2011kl, the light curves of 13 LFBOTs and the host galaxies of five ULGRBs from the literature in the context of the He-CO merger scenario. In Section~\ref{sec:data}, we describe the archival light curve data, the LFBOT model and our inference results; we present an alternative light curve model in the Appendix~\ref{app:fryer}. In Section~\ref{sec:host} we present an analysis of the host galaxies of our ULGRB sample. Finally, Section~\ref{sec:disc} contextualizes our findings. We assume a flat $\Lambda$CDM cosmology with $H_0 = 66.7\ \mathrm{km\ s^{-1}\ Mpc^{-1}}$ and $\Omega_M = 0.31$, consistent with \cite{aghanim2020planck}.

\section{LFBOT/ULGRB sample and Light Curve Modeling}
\label{sec:data}
\subsection{Data and Metzger Model Description}
GRB~111209A ($z\simeq0.677$; \citealt{2011GCN.12648....1V}) is the second longest known ULGRB with a prompt $T_\mathrm{90}\sim10^4$~s \citep{greiner2015very}. Its accompanying optical transient, SN~2011kl, is atypical of other GRB-SNe and has been extensively modeled (e.g., \citealt{cano2016self,ioka2016ultra,mazzali2016spectrum,wang2017modeling}), with most studies concluding that the transient light curve cannot be dominated by radioactive heating. We compile archival UV, optical and near-infrared (NIR) photometry for GRB~111209A/SN~2011kl from Swift/UVOT \citep{greiner2015very,ioka2016ultra}, Gemini and VLT \citep{levan2014new}, and GROND \citep{greiner2015very}. The underlying host-galaxy contributions in each filter and the host extinction ($E(B-V)_\mathrm{host}=0.04$) are taken from \citealt{greiner2015very}. 

For comparison, we collect archival UV through NIR photometry for the 11 LFBOTs studied in \cite{nugent_lfbots} in addition to MUSSES2020J \citep{jiang2022musses2020j} and AT~2024aehp \citep{sevilla2026multiwavelength}. These events are listed in Tab.~\ref{tab:lfbot_data}. For events with data extending $>80$ rest-frame days of coverage (AT~2018cow, AT~2024aehp and AT~2024wpp), we exclude observations beyond this phase to match with the modeling of \cite{metzger2015diversity}. AT~2022tsd notably exhibits minutes-duration flares at late times, which we exclude from our analysis \citep{ho2023minutes}. 

\begin{deluxetable*}{lcccc}
\tabletypesize{\scriptsize}
\tablecaption{LFBOT Comparison Sample\label{tab:lfbot_data}}
\tablehead{
\colhead{Event} &
\colhead{Redshift} &
\colhead{Filters} &
\colhead{Fitting window (days)} &
\colhead{Reference}
}
\startdata
CSS161010 & 0.033 & $UVW2,UVM2,UVW1,B,V,g,r,R,i,I,z$ & 80 & \citet{gutierrez2024css} \\
AT~2018lug & 0.271 & $g,r$ & 9 & \citet{ho2020koala} \\
AT~2018cow & 0.0141 & $UVW2,UVM2,UVW1,u,B,g,V,r,R,i,I$ & 80 & \citet{perley2019fast} \\
AT~2020mrf & 0.135 & $g,c,r,o$ & 36 & \citet{yao2022x} \\
AT~2020xnd & 0.244 & $UVW1,u,g,r,R,i,I,z$ & 28 & \citet{perley2021real} \\
MUSSES2020J & 1.063 & $g,r,i$ & 9 & \citet{jiang2022musses2020j} \\
AT~2022tsd & 0.256 & $g,r,i,o$ & 26 & \citet{ho2023minutes} \\
AT~2023fhn & 0.238 & $g,r,i$ & 10 & \citet{sevilla2026multiwavelength} \\
AT~2023hkw & 0.339 & $g,r,i,z$ & 10 & \citet{sevilla2026multiwavelength} \\
AT~2023vth & 0.0747 & $g,r$ & 17 & \citet{sevilla2026multiwavelength} \\
AT~2024qfm & 0.227 & $g,r,i$ & 35 & \citet{sevilla2026multiwavelength} \\
AT~2024aehp & 0.172 & $UVW2,UVM2,UVW1,u,g,r,i,z$ & 80 & \citet{sevilla2026multiwavelength} \\
AT~2024wpp & 0.0868 & $UVW2,UVM2,UVW1,u,B,g,V,c,r,R,o,i,I,z,J,H,K_s$ & 80 & \citet{lebaron2026most} \\
\enddata
\end{deluxetable*}

For both SN~2011kl and the LFBOT comparison sample, we fit the UVONIR data to a toy model presented in \cite{metzger2022luminous}; see their Section~2. In this framework, LFBOTs arise from the tidal disruption and hyper-accretion of a WR star onto a compact companion following a delayed merger. The optical emission is powered by (1) shock interaction between the material of the disrupted WR star and pre-existing CSM and (2) reprocessed radiation from the accretion onto the compact object or a jet. We let the following parameters vary: $M_{\rm *} \sim \mathcal{U}(5,100)\,M_\odot$, the mass of the WR star at time of merger; $M_{\rm pre}$, the pre-runaway envelope (or CSM) mass which controls shock heating deposition into the equatorial ejecta, which we parameterize as a fraction of $M_\mathrm{pre}=f_\mathrm{pre}\,M_*\sim \mathcal{U}(0.05,50)\,M_\odot$; $v_{\rm slow}\sim\mathcal{U}(0.1,10)$ kkm s$^{-1}$, the equatorial ejecta velocity;  $M_{\rm fast}\sim\mathcal{U}(0.1,10)\,M_\odot$, the fast polar ejecta mass along with its velocity $v_{\rm fast}\sim\mathcal{U}(0.01,0.5)\,c$; and $N_{\rm orbit}\sim\mathcal{U}(10,100)$, an effective number of binary orbits over which mass loss occurs (assuming a bare He core in Roche contact), which controls the radial extent of the pre-merger material and the timescale of the shock-heated emission. We introduce three additional free parameters.  $\gamma_{\rm UV}\sim\mathcal{U}(0,15)$ suppresses UV flux  at $\lambda_{\rm rest} \lesssim 3000$\,\AA\ following the form $F_\nu \propto (\lambda/3000\AA)^{\gamma_\mathrm{UV}}$. This is similar to the phenomenological model used for superluminous SNe (see e.g., \citealt{nicholl2017magnetar}) to account for Fe-line blanketing. $T_\mathrm{floor}\sim\mathcal{U}(1.5,30)$\,kK is a minimum temperature at which the assumed black body spectral energy distribution can reach before the photosphere begins to recede into the ejecta, typical in semi-analytic SN models (e.g., \citealt{guillochon2018mosfit}). Finally, $\sigma\sim\mathcal{U}(0,0.5)$ mag is a white noise term added to all measured uncertainties in quadrature to account for underestimation of observational errors and model mismatch. For all LFBOTs, we fit for time of merger; we take the time of merger to be the time of the GRB detection for SN~2011kl. For all events, we assume a fixed black hole mass $M_\bullet=10\,M_\odot$ for the compact binary companion. We explore an alternative model in Appendix~\ref{app:fryer}, which still arises from the He-CO merger scenario but attributes the first peak to shock cooling.

When modeling GRB~111209A/SN~2011kl, the separation between an afterglow-like vs thermal contribution depends, clearly, on the assumed afterglow properties. Interpretation of a late time break in GRB~111209A has been particularly contentious in the literature. \cite{greiner2015very} and \cite{kann2019highly} favor a steepening temporal index at $\sim9$ days, while \cite{ioka2016ultra} and \cite{gompertz2017magnetars} argue that such a break is not apparent in the X-ray (or optical) light curves. The lack of a break implies a wide jet opening angle but not unusually so.
Here, we simultaneously fit a standard synchrotron emission component with separable power laws in time and frequency:
\begin{equation}
F_{\nu}^{\rm AG}(t,\nu)
=
F_{\nu,0}
\left(\frac{t}{t_0}\right)^{-\alpha}
\left(\frac{\nu}{\nu_{\rm ref}}\right)^{-\beta},
\end{equation}
where $t_0$ and $\nu_\mathrm{ref}$ are constant, reference values not fit by the model. The AG component is added to the LFBOT model and a constant host galaxy per band (as reported in \citealt{greiner2015very}).


\subsection{Light Curve Fitting Results}

We fit all events using the Markov chain Monte Carlo inference package \texttt{emcee} \citep{foreman2013emcee}  with 100 walkers with 2000 steps each. Posterior medians and $1\sigma$ uncertainties are listed in Table~\ref{tab:lfbot_mcmc}. 

First, we discuss inferences for the broader LFBOT population. The WR mass $M_*$ is well-constrained for most events, where $M_*=M_\mathrm{fast}+M_\mathrm{slow}+M_\mathrm{acc}$ is the sum of the fast polar ejecta, the slower toroidal ejecta and the WR mass that accretes onto the BH companion (see Eqn. 15 of \citealt{metzger2022luminous}). Because we fix the BH companion mass (which ultimately impacts $M_\mathrm{acc}$), the exact values of $M_*$ should not be taken too literally. We find that the LFBOT population has a typical $M_*\sim20M_\odot$, with the notable exception of the high-mass AT~2022wpp with $M_*\sim40M_\odot$. 

There is substantial diversity across the population in the inferred masses and velocities for both the fast and slow ejecta components, although these parameters are often poorly constrained in our sample. Among events with reasonable constraints, there is a correlation between $M_\mathrm{pre}$ and $N_\mathrm{orb}$, in which events with a higher $M_\mathrm{pre}$ have a higher $N_\mathrm{orb}$ (AT~2020mrf, AT~2022tsd and AT~2024aehp), whereas events with a lower $M_\mathrm{pre}$ have a lower $N_\mathrm{orb}$ (CSS161010 and AT~2018cow). Similarly, events with a high $M_*$, and therefore higher $M_\mathrm{acc}$, have a higher $T_\mathrm{floor}$ (AT~2018lug, MUSSES2020J, AT~2022tsd, and AT~2023fhn, AT~2023vth, AT~2024wpp), whereas events with a lower $M_*$ have a lower $T_\mathrm{floor}$ (CSS161010, AT~2018cow, AT~2023hkw, AT~2024aehp and AT~2024qfm). Finally, we note that most events (due to a lack of rest-frame UV observations) do not have strong constraints on $\gamma_\mathrm{UV}$; for those that do, we find no evidence of UV flux suppression ($\gamma_\mathrm{UV}\lesssim0.5$).

\begin{figure}
\centering
\includegraphics[width=\linewidth]{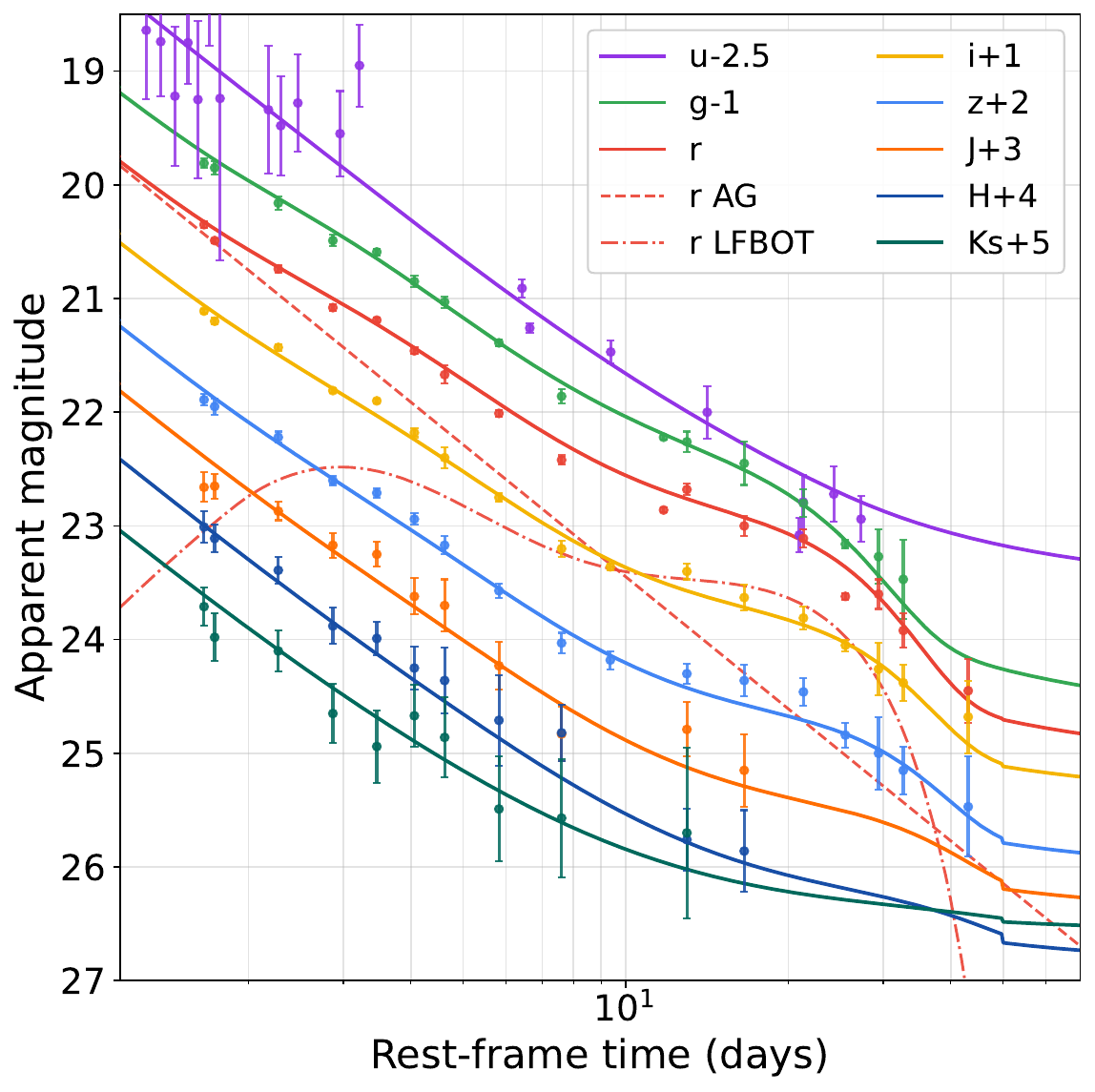}
\caption{Multi-band, joint fit for the GRB\,111209A afterglow and SN~2011kl optical transient. We use fixed host-galaxy AB magnitudes of $m_u = 26.0$, $m_g = 25.66$, $m_r = 25.04$, $m_i = 24.36$, $m_z = 24.02$, $m_J = 23.39$, $m_H = 22.84$, and $m_{K_s} = 21.56$ \citep{greiner2015very}.}
\label{fig:2011kl_lfbot-full-fit}
\end{figure}

We next turn to SN~2011kl, with fit parameters listed again in Table~\ref{tab:lfbot_mcmc}. As shown in Fig.~\ref{fig:2011kl_lfbot-full-fit}, we find a reasonable fit to UVONIR data, in which the LFBOT component becomes an apparent deviation from the AG $\sim3$ days post-merger. Compared to the LFBOT population, SN~2011kl has the lowest WR mass $M_*\simeq11M_\odot$ and, following a similar trend as the population, the lowest $T_\mathrm{floor}$. The number of orbits $N_\mathrm{orb}\simeq80$ over which mass loss occurs is larger than the majority of the LFBOT sample. While  $M_\mathrm{pre}\simeq1.9M_\odot, v_\mathrm{fast}\simeq0.2c$ and $v_\mathrm{slow}\simeq7$ km$^{-1}$ are all typical of the LFBOT population, the inferred polar mass, $M_\mathrm{fast}\simeq0.07M_\odot$ is notably low and only smaller than AT~2024aehp. The timescale of the early component $\tau\propto\sqrt{M/v}$, however, is typical of the LFBOT population. Finally, consistent with the spectrum of SN~2011kl (see Sec.~\ref{sec:disc}), we find strong evidence for UV suppression with $\gamma_\mathrm{UV}\simeq8$.

\begin{figure*}
\centering
\includegraphics[width=\linewidth]{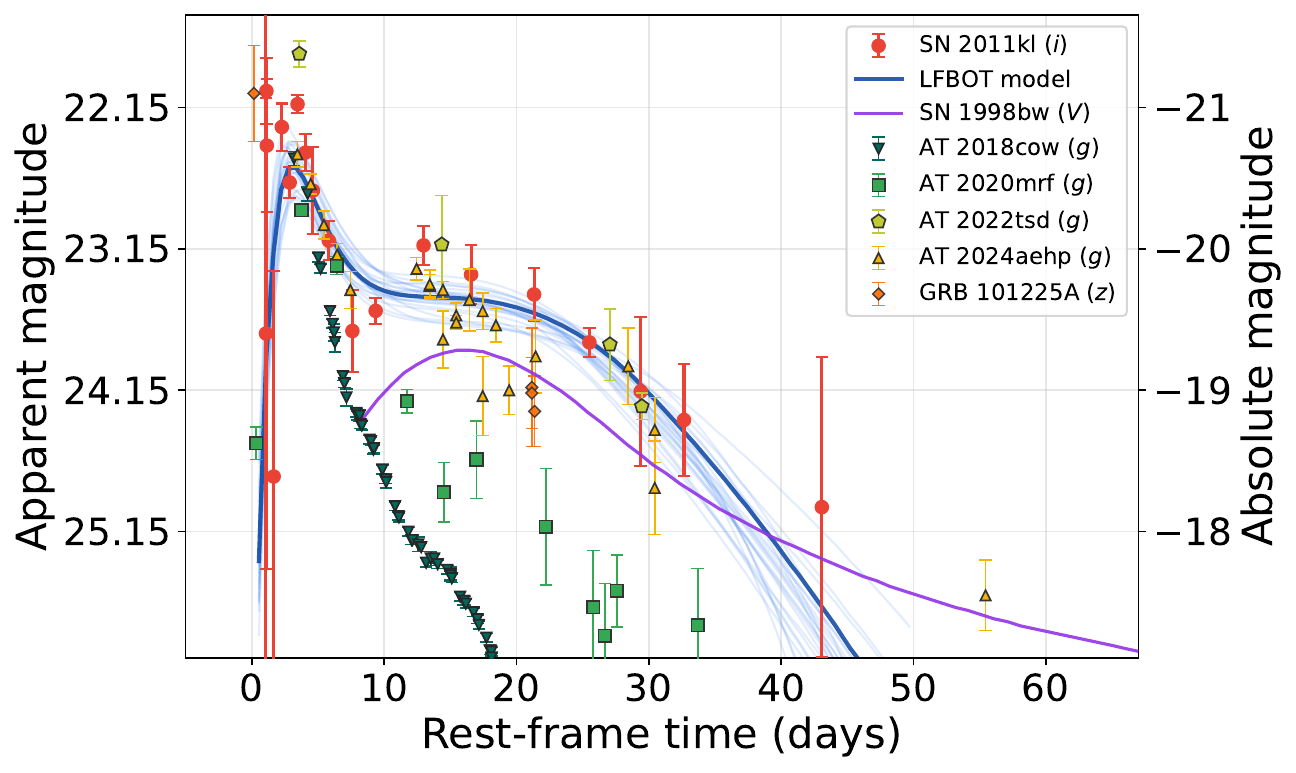}
\caption{
Rest-frame light curve of SN~2011kl (red circles, using $i$ as the a proxy for rest-frame $g$) with the fitted AG component removed compared to our best-fit LFBOT model (blue curves) and a sample of comparison transients. For reference, we show the light curve of the broad-lined Type Ic supernova SN~1998bw (purple curve; \citealt{clocchiatti2011ultimate}), as well as several LFBOTs: AT~2018cow (dark green triangles; \citealt{perley2019fast}), the somewhat longer-lived AT~2020mrf (green squares; \citealt{yao2022x}), AT~2022tsd (chartreuse pentagons; \citealt{ho2023minutes}), AT~2024aehp (yellow triangles; \citealt{sevilla2026multiwavelength}), and ULGRB GRB~101225A (orange diamonds, observer-frame $z$-band as proxy for rest-frame $g$; \citealt{levan2014new}). }
\label{fig:sample}
\end{figure*}

The rest-frame light curve of SN~2011kl (isolated from the AG using our best-fit model) is shown in Figure~\ref{fig:sample} along with select LFBOTs and GRB-SNe. At early times ($\lesssim10$ days), SN~2011kl closely tracks the rapid rise and high luminosity characteristic of most LFBOTs. However, its subsequent evolution is notably slower than that of the prototypical LFBOT AT~2018cow \citep{perley2019fast} and even the somewhat longer-duration AT~2020mrf \citep{yao2022x} and AT~2022tsd \citep{ho2023minutes}. Instead, SN~2011kl exhibits a more extended phase of luminous emission ($M\sim-19.5$, peaking at $\sim20$ days), with a shallower decline reminiscent of the plateau-like behavior observed in AT~2024aehp \citep{sevilla2026multiwavelength}.  In our model, this plateau can be attributed to higher mass loss pre-merger, either within the local environment ($M_\mathrm{pre})$ or codified as a large number of orbits.

While similar to some LFBOTs, SN~2011kl is shorter-lived, bluer and brighter than GRB-SNe, like the canonical SN~1998bw \citep{clocchiatti2011ultimate}, which has been noted many times in the literature (e.g., \citealt{levan2014new}). We additionally compare SN~2011kl to another well-sampled event ULGRB~101225A in Figure~\ref{fig:sample} \citep{greiner2015very}. We have not removed any afterglow contribution in this case, as the object appears to be immediately consistent with an expanding blackbody in the optical. Qualitatively, the two light curves are very similar, with both exhibiting an early peak at $M_g\simeq-21$ mag and a subsequent plateau until $\simeq20$ days. We briefly note that our LFBOT thermal model is unable to capture the incredibly rapid rise ($\lesssim1$ day) of GRB~101225A, so we do not present a formal fit.

\begin{figure*}[t]
  \centering
  \includegraphics[width=\textwidth]{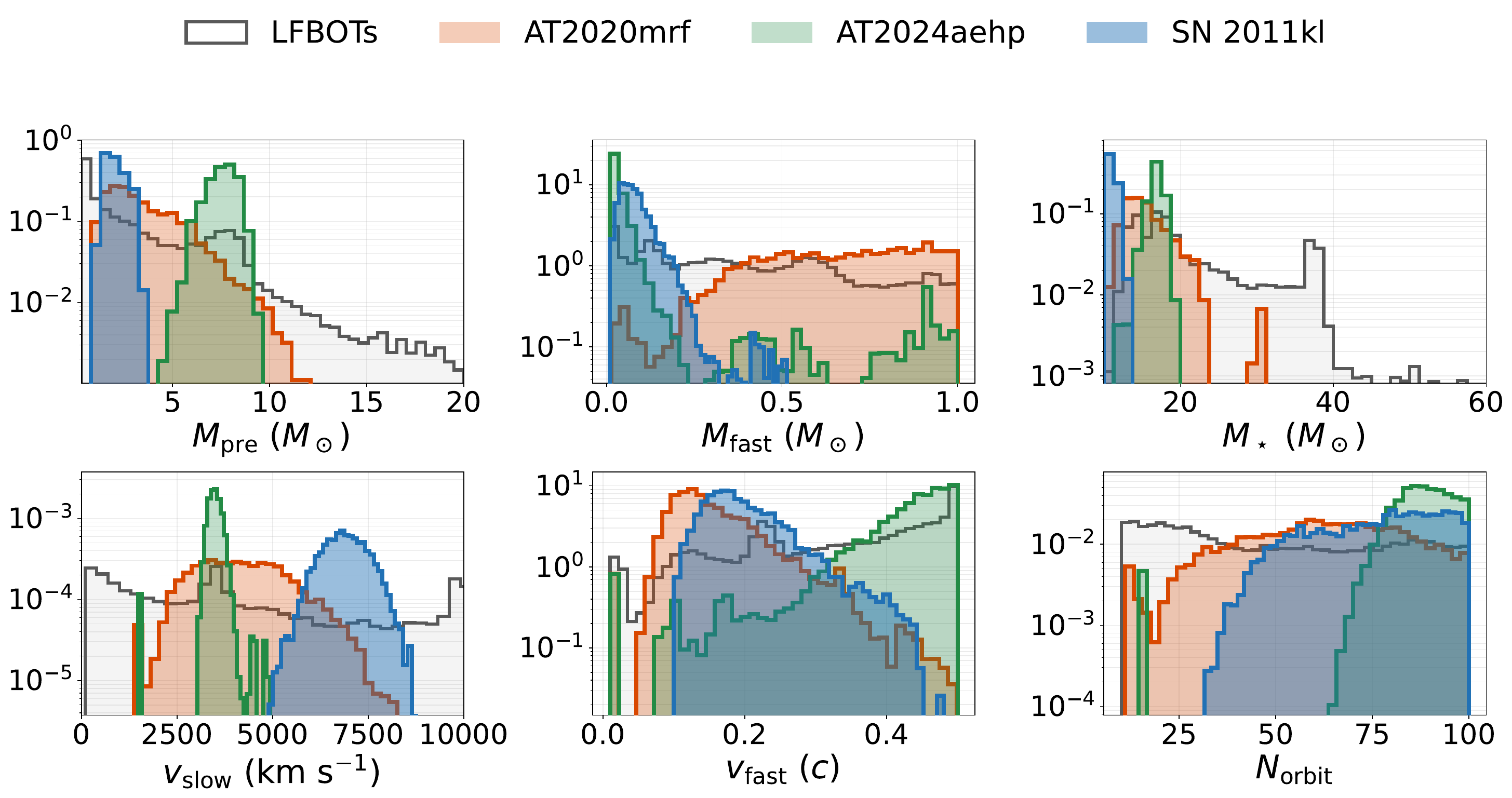}
  \caption{ LFBOT/SN~2011kl marginal posterior distributions for shared fit parameters.  Grey histograms show the joint LFBOT sample. Colored curves highlight longer-duration LFBOTs AT2020mrf and AT2024aehp which, similar to SN~2011kl, have higher inferred $N_\mathrm{orb}$ compared to the broader population. SN~2011kl (blue) is not included in the pooled LFBOT sample.}
  \label{fig:lfbot_posteriors}
\end{figure*}

\section{Host Galaxy Analysis}
\label{sec:host}
\subsection{Stellar Population Modeling}
\label{sec:sp_model}
As an additional point of comparison, we also contrast the host properties of ULGRBs and LFBOTs. Host galaxy stellar population properties (e.g., stellar mass, star formation rate, metallicity) have been crucial in defining the progenitors of various transient populations and, thus, inferring similarities between the ULGRB and LFBOT hosts may be further evidence of a shared progenitor origin. 

We consider five ULGRBs from the literature with measured redshifts: GRB~101225A (the ``Christmas burst"), GRB~111209A (associated with SN~2011kl), GRB~121027A, GRB~130925A and GRB~250702B (summarized in Table~\ref{tab:host}). The first three are the sample presented in \cite{levan2014new} that have sustained prompt emission over $\sim10^3$\,s and $L_\mathrm{iso}\gtrsim10^{49}$\,erg s$^{-1}$; GRB~130925A and GRB~250702B are the only other two archival events that also fit this definition.
We find host galaxy photometry for the ULGRB population in: \citet{thone2011unusual, levan2014new} (GRB 101225A), \citet{greiner2015very} (GRB 111209A), \citet{evans2014} (GRB 130925A), and \citet{carney2025} (GRB 250702B). For GRB 121027A, we collect archival host photometry in the Dark Energy Camera Legacy Survey Data Release 10 (DR10; \citealt{dey2019}). 

To model the stellar population properties of the ULGRB hosts and fit their aforementioned photometry, we use the Python-based stellar population modeling inference code, \texttt{Prospector}. \texttt{Prospector} determines properties such as total mass formed in the galaxy ($M_F$), the star formation history (SFH), stellar metallicity ($Z_*$), and optical depth of old ($\tau_{V,2}$) and young ($\tau_{V,1}$) stars. We employ a nested sampling fitting routine through \texttt{dynesty} to determine posterior distributions on the stellar population properties of interest and produce model SEDs with \texttt{FSPS} and \texttt{python-fsps}. Our \texttt{Prospector} model is nearly identical as the one used \citet{nugent_lfbots} to determine the properties of LFBOT hosts; thus, we refer the readers to that text for further details on the model specifications. The only change we make is in the form of the SFH; in \citet{nugent_lfbots}, LFBOT hosts are modeled with a seven-bin non-parametric SFH, given that the majority of their hosts have multiple photometric measurements and a high S/N spectrum. Here, all hosts only have four photometric filters (save for the host of GRB 101225A, which was only detected in two bands). Thus by employing a non-parametric SFH, we would be over-fitting the data. Instead, we choose a delayed-$\tau$ SFH, which has the functional form: $\text{SFH} \propto t*e^{-t/\tau}$, where $\tau$ is the e-folding factor and is a sampled parameter in the \texttt{Prospector} fit. This difference in SFH will have a subtle effect on the mass and present-day star formation rates (SFR) of the hosts: masses determined from non-parametric SFHs can be 25-100\% larger than those derived from parametric SFHs, and SFRs tend to slightly lower. Following the methods outlined in \citet{nugent2022}, we derive stellar masses ($M_{*,\mathrm{gal}}$), present-day SFRs and specific SFRs (sSFR = SFR/$M_{*,\mathrm{gal}}$), and mass-weighted ages ($t_m$) from the $M_F$ and SFH determined for each ULGRB host, and convert $\tau_{V,2}$ and $\tau_{V,1}$ to a total $V$-band magnitude ($A_V$). We present these derived properties in Table~\ref{tab:host}. Because the host of GRB 101225A only has two photometric points given its faintness, we can only robustly determine $M_{*,\mathrm{gal}}$, as all other properties require more photometric measurements to properly constrain. We further note that host stellar population properties for GRB111209A, GRB130925A, and GRB 250702B were determined in \citet{kann2018}, \citet{schady2015}, and \citet{carney2025}, respectively. Our stellar population property constraints are consistent within error bars to their work. Finally, we also find estimates of gas-phase oxygen abundance, $12 + \log(\textrm{O/H})$, in \citet{levan2014new} for the host of GRB111209A ($8.3\pm0.3$)  \citet{schady2015} for and the nearest star-forming region to GRB130925A ($8.70^{0.12}_{-0.15}$). Both studies used report $12 + \log(\textrm{O/H})$ using the $R_{23}$ method, based on detection of [OII]$\lambda\lambda3727,9$, [OIII]$\lambda\lambda4960,5007$, and H$\beta$ \citep{kobulinsky2004, kewley2019}. However, we emphasize that the spectra used for these estimates had contamination from their respective GRB afterglows, and thus were unusable in our \texttt{Prospector} fits.

\begin{figure*}
\centering
\includegraphics[width=\linewidth]{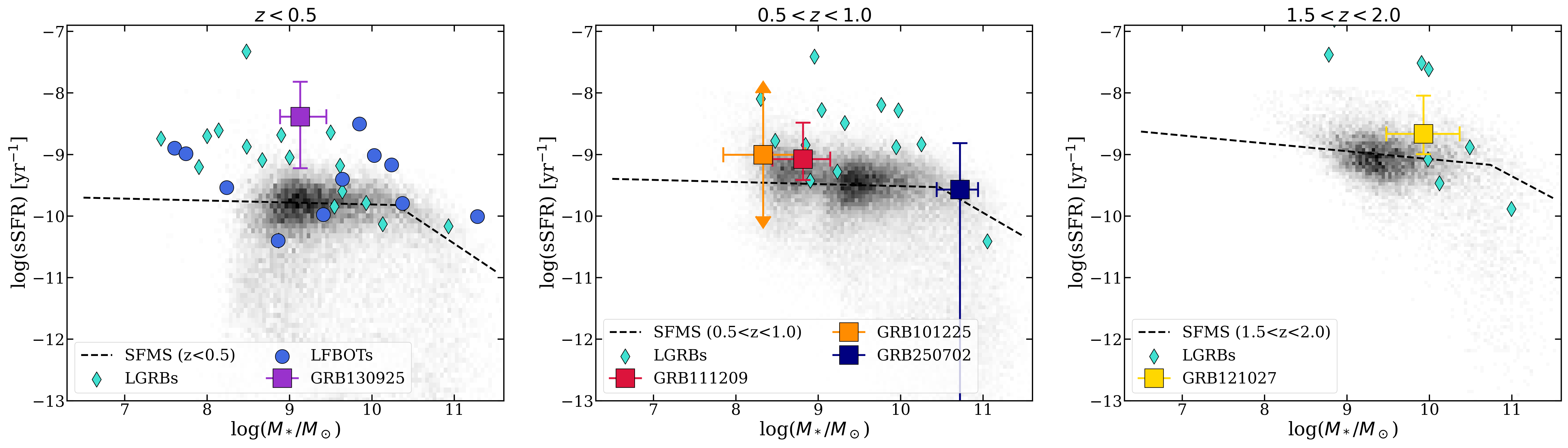}
\caption{The hosts of ULGRBs (multicolor squares), LFBOTs (blue circles), and LGRBs (cyan diamonds), compared to a field galaxy population from the 3D-HST and COSMOS2015 surveys \citep{skelton2014, laigle2016, leja2022}, separated into three redshift bins. We also highlight the SFMS determined in \citet{leja2022} for all three redshift bins (dashed black line). All three transient populations appear to be clustered above the SFMS, in lower mass ($<10^{10}$~$M_\odot$) galaxies with higher sSFRs than observed for the field galaxy populations. This indicates that these transients are closely tied to star formation for progenitor formation, and thus likely contain a young, massive stellar component within their progenitor systems.}
\label{fig:host}
\end{figure*}

\subsection{Comparison to LFBOT Hosts}
We next contrast the host properties of ULGRBs to those of 11 LFBOTs, studied in \citet{nugent_lfbots}. We begin by analyzing their sSFRs and stellar masses. Given that the current sample of ULGRBs ($0.35 \leq z \leq 1.77$) are observed at much higher redshifts than LFBOTs ($0.01 \leq z \leq 0.33$; \citealt{nugent_lfbots}), it is difficult to perform a direct comparison of the two host populations, as galaxy SFRs and stellar masses are known to evolve with redshift \citep{speagle2014, whitaker2014, leja2022}. Thus, we instead choose to compare the LFBOT and ULGRB host samples to field galaxy populations within specified redshift ranges as an indirect way of comparing their properties. 

In Figure~\ref{fig:host}, we plot ULGRB and LFBOT host sSFRs and stellar masses in three redshift bins ($z\leq0.5$, $0.5<z\leq1.0$, and $1.5 < z \leq 2.0$) against the COSMOS2015 \citep{laigle2016} and 3D-Hubble Space Telescope (HST; \citealt{skelton2014}) field galaxy populations presented in \citet{leja2022}. These galaxy stellar population properties were determined with a nearly identical \texttt{Prospector} model as employed in \citet{nugent_lfbots} for LFBOT hosts, and therefore are comparable to the properties studied here. We furthermore show the \citet{leja2022} star-forming sequence (SFMS), computed within each redshift bin, which is a well-understood galaxy relation that tracks how star-forming galaxies change in star formation as they gain in stellar mass. Lastly, as all LFBOTs are solely contained within the first redshift bin ($z<0.5$), we show an LGRB host population that extends throughout the full redshift range \citep{svensson2010, perley2013, wang2014, niino2017}. \citet{nugent_lfbots} showed that the individual distributions of stellar mass and sSFR of LFBOTs were not statistically different from those of LGRBs at $z<0.5$ (although their metallicity distributions were distinct), thus LGRBs may be a good proxy transient host population for understanding how LFBOT hosts evolve at higher-$z$.

While the population of ULGRBs and their hosts is quite small, and any concrete conclusions on their host properties is difficult to make, we do find similarities between their hosts and those of LFBOTs and LGRBs. First, it appears that the majority of all three transients' hosts lie above the SFMS, distinct from the bulk of the field galaxy population. We also find that the majority of ULGRB hosts, similar to LFBOT and LGRB hosts, are lower stellar mass ($<10^{10}$~$M_\odot$).  Similar to conclusions made for LFBOT and LGRB hosts, this likely suggests that the ULGRB progenitor is strongly tied to star formation and possibly comprises a young, massive star (commensurate with conclusions made from observations of the transients themselves; \citealt{thone2011unusual, neights2025fermi}). Moreover, the lack of distinction between ULGRB host $M_*$ and sSFR to those of LFBOTs and LGRBs implies that we cannot rule out a common origin for all three transients from host galaxies alone. 

We finally turn our attention to the $12+\log(\textrm{O/H})$ metallicities of ULGRB hosts, compared to those of LFBOTs. \citet{nugent_lfbots} determined that LFBOTs have population median (and 68\% interval) $12+\log(\textrm{O/H}) = 8.59^{+0.18}_{-0.24}$. The two ULGRBs with metallicity measurements (GRBs 111209A and 130925A) both fall within this range. However, we once again note that it is difficult to make conclusive remarks with only two events, both of which are at higher redshifts than any LFBOTs, where we expect metallicity to decrease. Thus, while we cannot validate that this is evidence that ULGRBs and LFBOTs have a shared origin, we also cannot rule out the possibility that they do from their environmental properties.


\section{Discussion and Conclusions}
\label{sec:disc}

Here, we review key insights from the light curve and host modeling of SN~2011kl vs the LFBOT population.
We have shown that a simple LFBOT model is consistent with the thermal, optical transient SN~2011kl. However, the optical counterpart to SN~2011kl differs in important ways from typical LFBOTs. In the first week, SN~2011kl is quite similar to canonical LFBOTs like AT~2018cow; however, the light curve then exhibits a longer ``plateau" phase ($\simeq20$\,days). Within our modeling framework of a He+CO binary, this extended emission can be explained by either more orbits over which mass loss occurs or a larger equatorial mass reservoir surrounding the central engine (see Fig.~3 of \citealt{metzger2022luminous}). 

LFBOTs AT~2020mrf, AT~2022tsd and AT~2024aehp similarly have large $N_\mathrm{orb}$ (although the inferences for AT~2020mrf and AT~2022tsd are more marginal). Interestingly, AT~2020mrf and AT~2022tsd are X-ray bright events, reaching $\sim10^{43}-10^{44}$ erg s$^{-1}$ approximately $\sim20$ days post-explosion; AT~2024aehp has limits $L_X<10^{43.3}$ erg s$^{-1}$ at earlier times $t\lesssim15$\,days. X-ray observations of SN~2011kl extend to $\sim10$ days rest-frame after the GRB, at which point the source remains luminous $(\sim10^{44}$ erg s$^{-1}$) and consistent with simple AG decay \citep{levan2014new,margutti2019embedded}. However, given the scatter in late-time measurements, a component similar to X-ray bright LFBOTs cannot be ruled out. AT~2024aehp is also unusual for an LFBOT in the radio, brightening over 100 days after discovery. From these observations, \cite{sevilla2026multiwavelength} estimate $\dot{M}\sim3\times10^{-4}M_\odot/\mathrm{yr}$ from a synchrotron self-absorption model. While derived under different assumptions, radio afterglow modeling of GRB~111209A/SN~2011kl yields an unusually high local density with $A_*\gtrsim2-70$ \citep{gompertz2017magnetars}, corresponding to $\dot{M}\sim(0.2-7)\times10^{-4}M_\odot/\mathrm{yr}$, comparable to the value inferred for AT~2024aehp. We note that the delayed radio brightening for AT~2024aehp may also be suggestive of an off-axis jetted origin. 

Compared to the LFBOT sample, SN~2011kl has the lowest inferred stellar mass at the time of merger $M_*\simeq11M_\odot$. 
If this relative low mass reflects a relatively lower-mass progenitor, it may be consistent with the scenario explored by \cite{fryer2025explaining}, in which lower-mass WR/He stars produce lower accretion rates and consequently longer-lived, high-angular momentum accretion disks capable of powering ULGRBs. Alternatively, the low inferred mass may instead indicate a progenitor system more heavily stripped of He. This would be consistent with our large inferred $N_\mathrm{orb}$, which suggests a longer phase of interaction between the compact object and progenitor prior to merger. 

We also find that SN~2011kl's light curve exhibits strong UV suppression at wavelengths $\lambda \lesssim 3000$~\AA, consistent with its observed spectrum \citep{greiner2015very,mazzali2016spectrum}. Our inferred suppression parameter ($\gamma_\mathrm{UV}\simeq 9$) is significantly larger than typical values for SLSNe, which are generally $\simeq2$, although with a considerable spread (see \citealt{gomez2024type}; note that their prior excludes $\gamma_\mathrm{UV}>5$). This is also atypical for LFBOTs, which are typically UV bright when such observations exist; however, similar to LFBOTs, the optical spectrum of SN~2011kl is largely featureless beyond 3000\AA. This has previously been explained by a high velocity and not an abnormally high metallicity (in fact, a subsolar metallicity is inferred; \citealt{greiner2015very,mazzali2016spectrum}). \cite{aspegren2026emission} recently explored the suppression of UV/optical lines in LFBOT-like transients. They find that it is particularly challenging to suppress optical C II, Ca II and O II lines in H-/He-poor events,  supporting the notion that high velocities are likely a requirement to explain the absence of optical features in SN~2011kl. 

It is unclear why SN~2011kl has more UV suppression than any of the LFBOTs with rest-frame UV observations but less suppression than typical GRB-SNe. It may reflect progenitor metallicity (with SN~2011kl being more metal-rich than other LFBOTs). However, this is not necessarily consistent with the global host galaxy metallicity. SN~2011kl has a somewhat subsolar inferred stellar metallicity ($\log(Z_*/Z_\odot)\simeq-0.5$), very typical of the LFBOT population ($\log(Z_*/Z_\odot)\sim-1.5$ to $+0.1$; \citealt{nugent_lfbots}). Alternatively, it is possible that local dust attenuation contributes to the UV suppression, although unlikely. For SN~2011kl, we expect the earliest mass loss during runaway to be at a radius of $R\sim10^{14}$ cm (see Eqn.~30 of \citealt{metzger2022luminous}); however, the dust sublimation radius of the transient would likely be beyond this \citep{metzger2023dust}. Any dust associated with this material would therefore be destroyed by the UV emission of the transient itself. Finally, it is possible that the relative contribution of multiple geometric components contribute to the SED. SN~2011kl's light curve may be dominated by a more extended emitting region than that of the LFBOT population, either due to intrinsic differences in the outflow structure or viewing-angle effects \citep{aspegren2026emission}. Such effects may be more apparent with dedicated spectral sequences. Unfortunately, the only available spectrum for SN~2011kl was taken $\simeq9$ days post-merger, just before LFBOT AT~2018cow began to show spectroscopic evidence for H-rich CSM.

Next, we turn to inferred properties of the jet. Our best-fit model does not require a jet break (consistent with the X-ray photometry; \citealt{ioka2016ultra,gompertz2017magnetars}), suggesting a somewhat wide ($\theta>10^{\circ})$ jet opening angle. To allow for a SN-like optical component, both \cite{greiner2015very} and \cite{kann2019highly} require a jet break at $\sim9$ days post-merger, despite a lack of such a break in the X-ray or UV light curves. More broadly, ULGRBs seem to have a wide range of possible jet opening angles (see e.g., the narrow jet inferred for the more recent GRB~250702B; \citealt{o2025comprehensive}). In contrast, no LFBOT has been associated with a successful, ultra-relativistic jet, despite clear evidence for central engines and mildly relativistic outflows (e.g., \citealt{margutti2019embedded,coppejans2020mildly,ho2023minutes,aj2025most}).  

We further show that the host properties of ULGRBs (stellar masses, sSFRs, and metallicities) are not clearly distinct from those of LFBOTs and LGRBs, both of which may also arise from compact object stellar mergers \citep{fryer1998helium, metzger2022luminous, klencki2025luminous}. All three transient populations appear to occur in galaxies with higher sSFRs and lower stellar masses than the bulk of field galaxy populations at similar redshifts. This likely shows that ULGRBs, similar to LFBOTs and LGRBs, are strongly tied to star formation for progenitor formation and, thus, likely have a young, stellar component in their progenitor system. The He+CO merger scenario may also naturally favor sub-solar
metallicities, although the origin of this preference likely depends on
the specific evolutionary channel. In the delayed dynamical instability
scenario \citep{klencki2025luminous}, lower-metallicity stars are more
compact, increasing the parameter space for binaries that avoid
interaction until after the donor has evolved off the main sequence and
developed a helium core (e.g., \citealt{klencki2020massive}). For He+CO mergers
occurring in post-CE systems, as in \cite{metzger2022luminous}, sub-solar
metallicity may instead favor the re-expansion of stripped He stars
\citep{laplace2020expansion}, which could provide the trigger for the final
He+CO merger \citep{fryer2025explaining}. Finally, if the CO is a BH, an
additional low-Z preference may arise because weaker winds in metal-poor
stars favor BH formation, although this effect is not expected to
dominate for the relatively low-mass BHs that ULGRBs may prefer \citep{fryer2025explaining}.

We reiterate that the population of ULGRBs is still currently too small to validate based on host properties whether this is indicative that ULGRBs, LFBOTs, and LGRBs share the same or similar progenitors. Moreover, increasing the sample size of ULGRBs, even minimally, may push the host  properties further from those of LFBOTs and LGRBs. Thus, future studies on expanded populations of ULGRBs should continue host comparisons between these transient samples to deduce whether they do or do not indicate a possible shared origin. 

We briefly highlight two correlations within the broader LFBOT sample that are consistent with the He-CO merger picture. For well-observed events, we find that LFBOTs with larger CSM reservoirs also have larger values of $N_\mathrm{orb}$. The interpretation of this trend is subtle because $N_\mathrm{orb}$ in our model is defined in the context of a bare He core in Roche contact. It is possible that, instead, unstable mass transfer begins while the donor retains a more extended envelope (i.e., the scenario explored in \citealt{klencki2025luminous}). In this case, the merger is immediately preceded by the dynamical inspiral of the BH/CO through the remaining H-envelope of the star. The orbital period at merger can be substantially longer, and even a modest number of orbits may correspond to a large ``effective" $N_\mathrm{orb}$ in our model framework. Thus, the observed $M_\mathrm{pre}-N_\mathrm{orb}$ correlation may reflect a continuum in the degree of envelope stripping: less-stripped systems provide both a larger CSM reservoir and longer orbital periods over which to lose said mass. Note that the circumstellar density profile, however, may differ substantially between stripped-He and partially stripped merger scenarios. The CSM in the former case is driven by runaway L2 mass loss prior to merger, while the latter is shaped by the extended stellar envelope. From the small sample, it is unclear if this trend represents a continuum of sources or two distinct merger pathways. Interestingly, SN~2011kl does not follow this trend. We infer a fairly average $M_\mathrm{pre}$, despite a large $N_\mathrm{orb}$. This may indicate that SN~2011kl is extremely stripped compared to the LFBOT population, consistent with other classical GRB-SNe.  We additionally find evidence that LFBOTs with larger stellar masses at the time of merger remain hotter during their evolution (i.e., have a larger $T_\mathrm{floor}$); SN~2011kl does follow this trend. In our framework the mass accreted onto the BH scales with the mass of the WR/He star, $M_\mathrm{acc}\propto M_*^{0.65}$. This may imply that a larger accreted mass is leading to higher/more sustained engine power and ultimately hotter ejecta.

Comparing the volumetric rates and redshift evolution of ULGRBs and LFBOTs can help determine whether ULGRBs are a subset of the broader LFBOT population. Although highly uncertain, the rates are consistent with one another. \cite{ho2023search} find that AT~2018cow-like LFBOTs are, at most, 0.01\% of the local CCSN rate ($0.3-420$ yr$^{-1}$ Gpc${-3}$). Using \textit{Swift}, the volumetric ULGRB progenitors has been (roughly) estimated as $\sim$1-30~Gpc$^{-3}$~yr$^{-1}$ (depending on the typical beaming correction; \citealt{prajs2017volumetric}), or $\sim$2\% of the BAT GRB population \citep{lien2016third}. 
However, given the limitations of field of view and interruption of coverage for \textit{Swift} due to Earth, this is likely also a lower limit. 
Making progress requires a more sensitive detector far from Earth, with continuous and contiguous observations over $\sim$days timescales, ideally with precise localizations. Instruments like Konus-\textit{Wind} and \textit{Psyche}-GRNS are therefore essential to identify and fully characterize these events \citep{neights2025fermi}. Surveys like the Legacy Survey of Space and Time (LSST; \citealt{ivezic2019lsst}) and Argus \citep{law2022low} should dramatically increase the discovery rate of LFBOTs, if they can be identified in time for spectroscopic confirmation. A larger and less biased sample of ULGRBs and LFBOTs will be essential for understanding whether these populations occupy a common continuum of engine-driven transients. 

Last, we outline other observational predictions if ULGRBs and LFBOTs share a common progenitor channel. First, future ULGRBs should exhibit LFBOT-like optical counterparts, potentially followed by luminous, plateau-like emission (similar to AT~2024aehp and SN~2011kl). UV through NIR coverage in the days to weeks following ULGRBs is critical for modeling such behavior. Similarly, at least a subset of LFBOTs are expected to host successful, off-axis jets. Long term radio followup (e.g., with the Deep Synoptic Array 2000) can constrain energetic GRBs even up to $\sim90^\circ$ off axis \citep{schroeder2025late}. Radio observations of on-axis ULGRBs, particularly to $t\gtrsim200$ days when possible, provide another test for a shared progenitor. Many LFBOTs exhibit luminous radio emission powered by interaction with dense CSM extending to $R_\mathrm{CSM}\sim\mathrm{few}\times10^{16}$ cm. Similar excess radio emission should be visible in ULGRBs if their progenitors have similar geometries, although disentangling the AG and LFBOT-like components may be challenging for on-axis events. Likewise, dust associated with the extended CSM may produce delayed infrared emission or dust echoes similar to those proposed for AT~2018cow \citep{metzger2023dust}. Mid-infrared observations with the \textit{James Webb Space Telescope} weeks to months after an ULGRB may provide a clearer picture of the circumstellar environment. It is also the case that, if ULGRBs arise primarily/exclusively from the He-CO merger scenario, we expect minimal newly formed $^{56}$Ni \citep{2019MNRAS.488..259F}, unlike GRB-SNe. Late time, rest-frame optical followup into the nebular phase can test this prediction. Finally, we expect a broader diversity in host galaxy metallicities of ULGRBs than is observed for classical LGRBs and SLSNe if ULGRBs and LFBOTs arise from lower-mass BH or NS-forming binary systems rather than the most massive collapsar progenitor. A larger sample of ULGRBs may unveil a continuum in host properties spanning LFBOTs, ULGRBs, and classic LGRBs.

\begin{acknowledgments}
We thank I.~Andreoni, O.~Gottlieb, E.~Quataert, and D.~Tsuna for useful discussions. The Villar Astro Time Lab acknowledges support through the David and Lucile Packard Foundation, the Research Corporation for Scientific Advancement (through a Cottrell Fellowship and Bridge Award) and the National Science Foundation under AST-2433718, AST-2407922 and AST-2406110. This work is supported
by the National Science Foundation under Cooperative Agreement PHY-2019786 (the NSF AI Institute for
Artificial Intelligence and Fundamental Interactions). Analysis presented here includes code written using Cursor.

BDM gratefully acknowledges support from the National Science Foundation (grant AST-2406637) and NASA (grants 80NSSC24K0408, 80NSSC26K0299, 80NSSC22K0807).  The Flatiron Institute is supported by the Simons Foundation.  

The work by CLF was supported by the U.S. Department of Energy, Office of Science, Office of Advanced Scientific Computing Research, Department of Energy Computational Science Graduate Fellowship under Award Number DE-SC0024386. 

This work is partially based on observations obtained with the Samuel Oschin Telescope 48-inch and the 60-inch Telescope at the Palomar Observatory as part of the Zwicky Transient Facility project. ZTF is supported by the National Science Foundation and a collaboration of partners, with operations conducted by COO, IPAC, and UW. We additionally acknowledge the use of public data from the Swift data archive. Based in part on data collected at the Subaru Telescope and retrieved from the HSC data archive system, which is operated by Subaru Telescope and Astronomy Data Center at National Astronomical Observatory
of Japan. The Hyper Suprime-Cam (HSC) collaboration includes the astronomical communities of Japan
and Taiwan, and Princeton University. The HSC instrumentation and software were developed by the National Astronomical Observatory of Japan (NAOJ), the
Kavli Institute for the Physics and Mathematics of the
Universe (Kavli IPMU), the University of Tokyo, the
High Energy Accelerator Research Organization (KEK),
the Academia Sinica Institute for Astronomy and Astrophysics in Taiwan (ASIAA), and Princeton University. Funding was contributed by the FIRST program
from Japanese Cabinet Office, the Ministry of Education, Culture, Sports, Science and Technology (MEXT),
the Japan Society for the Promotion of Science (JSPS),
Japan Science and Technology Agency (JST), the Toray
Science Foundation, NAOJ, Kavli IPMU, KEK, ASIAA,
and Princeton University.

Based on observations obtained at the international Gemini Observatory, a program of NSF NOIRLab, which is managed by the Association of Universities for Research in Astronomy (AURA) under a cooperative agreement with the U.S. National Science Foundation on behalf of the Gemini Observatory partnership: the U.S. National Science Foundation (United States), National Research Council (Canada), Agencia Nacional de Investigación y Desarrollo (Chile), Ministerio de Ciencia, Tecnología e Innovación (Argentina), Ministério da Ciência, Tecnologia, Inovações e Comunicações (Brazil), and Korea Astronomy and Space Science Institute (Republic of Korea).

This research is based on observations made with the NASA/ESA Hubble Space Telescope obtained from the Space Telescope Science Institute, which is operated by the Association of Universities for Research in Astronomy, Inc., under NASA contract NAS 5–26555. 

This project used data obtained with the Dark Energy Camera (DECam), which was constructed by the Dark Energy Survey (DES) collaboration. Funding for the DES Projects has been provided by the U.S. Department of Energy, the U.S. National Science Foundation, the Ministry of Science and Education of Spain, the Science and Technology Facilities Council of the United Kingdom, the Higher Education Funding Council for England, the National Center for Supercomputing Applications at the University of Illinois at Urbana-Champaign, the Kavli Institute of Cosmological Physics at the University of Chicago, Center for Cosmology and Astro-Particle Physics at the Ohio State University, the Mitchell Institute for Fundamental Physics and Astronomy at Texas A\&M University, Financiadora de Estudos e Projetos, Fundacao Carlos Chagas Filho de Amparo, Financiadora de Estudos e Projetos, Fundacao Carlos Chagas Filho de Amparo a Pesquisa do Estado do Rio de Janeiro, Conselho Nacional de Desenvolvimento Cientifico e Tecnologico and the Ministerio da Ciencia, Tecnologia e Inovacao, the Deutsche Forschungsgemeinschaft and the Collaborating Institutions in the Dark Energy Survey. The Collaborating Institutions are Argonne National Laboratory, the University of California at Santa Cruz, the University of Cambridge, Centro de Investigaciones Energeticas, Medioambientales y Tecnologicas-Madrid, the University of Chicago, University College London, the DES-Brazil Consortium, the University of Edinburgh, the Eidgenossische Technische Hochschule (ETH) Zurich, Fermi National Accelerator Laboratory, the University of Illinois at Urbana-Champaign, the Institut de Ciencies de l’Espai (IEEC/CSIC), the Institut de Fisica d’Altes Energies, Lawrence Berkeley National Laboratory, the Ludwig Maximilians Universitat Munchen and the associated Excellence Cluster Universe, the University of Michigan, NSF’s NOIRLab, the University of Nottingham, the Ohio State University, the University of Pennsylvania, the University of Portsmouth, SLAC National Accelerator Laboratory, Stanford University, the University of Sussex, and Texas A\&M University.

BASS is a key project of the Telescope Access Program (TAP), which has been funded by the National Astronomical Observatories of China, the Chinese Academy of Sciences (the Strategic Priority Research Program “The Emergence of Cosmological Structures” Grant \# XDB09000000), and the Special Fund for Astronomy from the Ministry of Finance. The BASS is also supported by the External Cooperation Program of Chinese Academy of Sciences (Grant \# 114A11KYSB20160057), and Chinese National Natural Science Foundation (Grant \# 12120101003, \# 11433005).

The Legacy Survey team makes use of data products from the Near-Earth Object Wide-field Infrared Survey Explorer (NEOWISE), which is a project of the Jet Propulsion Laboratory/California Institute of Technology. NEOWISE is funded by the National Aeronautics and Space Administration.

The Legacy Surveys imaging of the DESI footprint is supported by the Director, Office of Science, Office of High Energy Physics of the U.S. Department of Energy under Contract No. DE-AC02-05CH1123, by the National Energy Research Scientific Computing Center, a DOE Office of Science User Facility under the same contract; and by the U.S. National Science Foundation, Division of Astronomical Sciences under Contract No. AST-0950945 to NOAO.

This work further utilized observations from the Gran Telescopio Canarias (GTC)/La Palma and the La Silla Observatory's Gamma-ray Burst Optical/Near-infrared Detector (GROND).

Software citation information aggregated using \texttt{\href{https://www.tomwagg.com/software-citation-station/}{The Software Citation Station}} \citep{software-citation-station-paper,software-citation-station-zenodo}.

\end{acknowledgments}

\appendix
\section{Alternative Light Curve Model}
\label{app:fryer}
We have also fit the data using a two-component shock heating model using the SN light-curve code, SNLC~\citep{niblett2025studying}.  SNLC employs a gray flux-limited diffusion transport scheme using line-binned opacities from the LANL atomic database team~\citep{colgan2016new}.  This code includes a number of heating schemes:  energy from radioactive decay (including a leakage scheme for the gamma-ray transport), central engine power sources (magnetar or fallback), and shock heating.  

Our base progenitor is a 7\,M$_\odot$ Wolf-Rayet star artificially exploding using an energy deposition power source~\citep{fryer2018parameterizing}.  The SN shock is followed until it reaches the edge of the star.  We then alter this base progenitor to produce a range of explosion properties varying the stellar radius, the ejecta mass and velocity.  For this exploratory work, we do not vary the stellar radius or ejecta mass beyond this basic progenitor.  However, we increase the velocity by a factor of 2.5, corresponding to an isotropic explosion energy of $4 \times 10^{52} \, {\rm erg}$. This is higher than the inferred kinetic energy from the \cite{metzger2022luminous} model ($\sim10^{52}$\,erg). The WR progenitor mass, however, is similar to the inferred stellar mass in our \cite{metzger2022luminous} model ($M_*\sim10\,M_\odot$).

We also vary the heating schemes:  we can tune the $^{56}Ni$ mass and distribution, magnetar spins and magnetic fields, accretion energy and shock heating.  For this paper, we focus on shock heating methods only:  $^{56}Ni$ mass is set to zero and no internal energy (magnetar or accretion sources) are included, consistent with the low $^{56}Ni$ masses found in e.g., \cite{2019MNRAS.488..259F}.  The other sources tend to make broader light curves~\citep{niblett2025studying} because the energy must diffuse to the photosphere before they can power the observed light curve.

\begin{figure}[!htb]
  \centering
  \includegraphics[width=\linewidth]{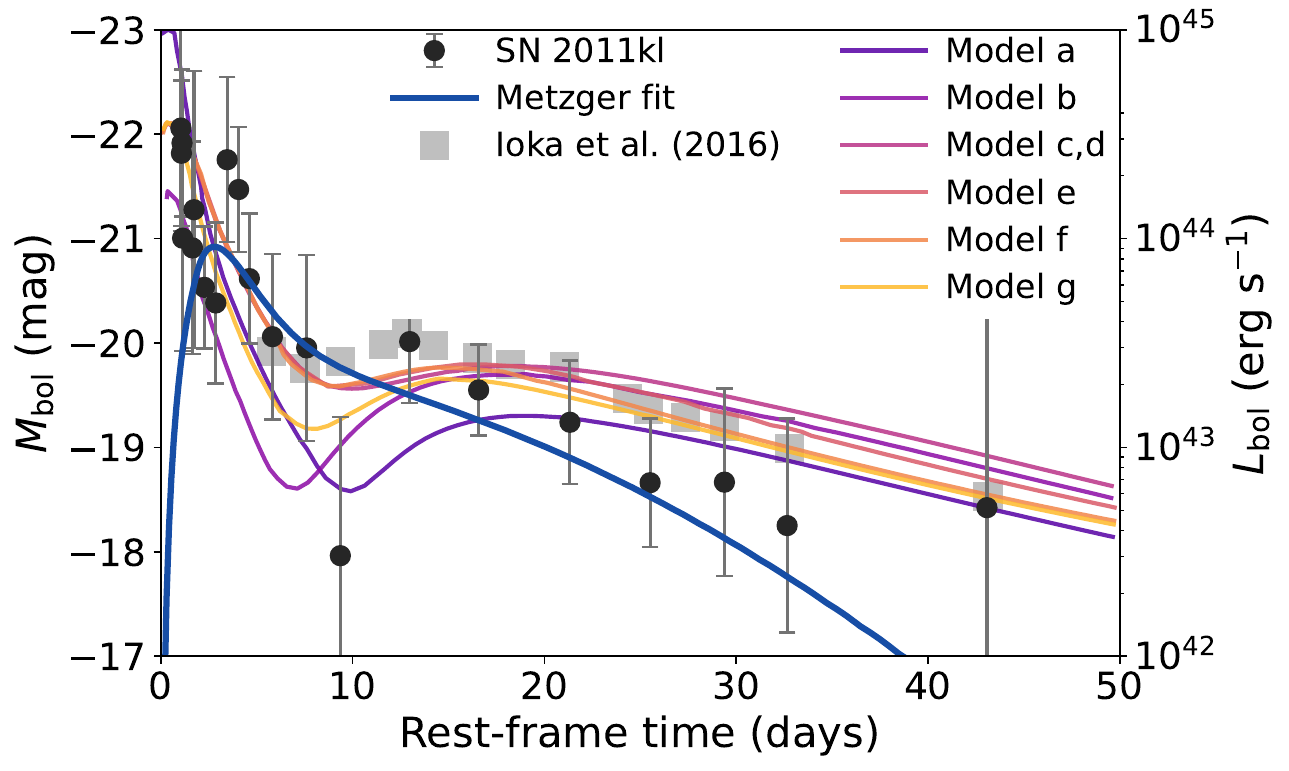}
  \caption{Bolometric light curve of SN~2011kl and comparative models in the rest frame.  Black points show our pseudo-bolometric estimates for SN~2011kl. For comparison, we show the bolometric estimate from \cite{ioka2016ultra}, which is in broad agreement with our estimate. 
  Colored curves show SNLC shock-heating models (Models~a--g); model parameters are given in Table~\ref{tab:snlc_models}.
  Models~c and~d differ only in outer shell radius and are shown as a single (overlapping) curve.}
  \label{fig:bol_2011kl_overlay}
\end{figure}

We consider a two-component model:  initial shock breakout and cooling, shock interactions with the interstellar medium. The model parameters explored are listed in Table~\ref{tab:snlc_models} and shown along with our bolometric light curve in Fig.~\ref{fig:bol_2011kl_overlay}. For the initial shock breakout, we seek to mimic shock interactions in the dense WR, stellar wind~\citep[e.g.][]{fryer2020role}.  For the dense stellar winds of WR stars, shock breakout in a WR star does not occur at the edge of the star but in the inner stellar wind.  Here we assume the kinetic energy of the shock is deposited into the outer layers of the star.  We vary both the fraction of the kinetic energy converted into heat (1-5\% $- f_{SBOE}$) and the fraction of the ejecta mass (moving from the outer ejecta inward) heated as the shock propagates through the inhomogeneous wind (1-5\% $- f_{SBOM}$).  This powers the early time light-curves as the blastwave breaks out of the wind and subsequently cools during the envelope cooling phase. 

For our second component, shock interaction with the CSM, we employ the method described in \cite{niblett2025studying}.  For these calculations, we assume the blast wave is striking a shell of material ejected in a common envelope phase.  As it propagates through this material, it decelerates, depositing its momentum and energy into the shell.  Here we can vary the mass of the shell ($m_{\rm shell}$), its inner ($r_{\rm shell}^{\rm in}$)and outer radii ($r_{\rm shell}^{\rm out}$).  With these 5 parameters, we produce a series of light curves to determine how well a SBO/cooling plus shell interaction model fits the current data.  Figure~\ref{fig:bol_2011kl_overlay} compares the bolometric light curves for a subset of our simulations against the observations.  The best fit models use roughly 1\% of the energy from the shock deposited in the outer 1-2\% of the star to match the early (first ~5 days).  The second component, shell interaction, required a 1.5-4\,M$_\odot$ shell spanning $(0.5-1)\times10^{14}$\,cm to $(2.5-4)\times10^{15}$\,cm. This shell mass is comparable to the mass $M_\mathrm{pre}\sim2\,M_\odot$ inferred for the \cite{metzger2022luminous} model. The shell extent is similarly consistent with the expectations from the runaway mass loss. 

\clearpage
\movetabledown=0.5in
\begin{longrotatetable}
\begin{deluxetable}{lccccccccccc}{}
\tabletypesize{\scriptsize}
\tablewidth{0pt}
\tablecaption{SN~2011kl/LFBOT posterior medians with 16th--84th percentile intervals.\label{tab:lfbot_mcmc}}
\tablehead{
\colhead{Event} & \colhead{$z$} & \colhead{$M_{\rm pre}$} & \colhead{$N_{\rm orb}$} & \colhead{$v_{\rm slow}$} & \colhead{$M_{\rm fast}$} & \colhead{$v_{\rm fast}$} & \colhead{$M_\star$} & \colhead{$M_{\rm acc}$\tablenotemark{a}} & \colhead{$\alpha_{\rm AG}$} & \colhead{$\beta$} & \colhead{$m_{0,\rm ref}$} \\
\colhead{} & \colhead{} & \colhead{($M_\odot$)} & \colhead{} & \colhead{(km\,s$^{-1}$)} & \colhead{($M_\odot$)} & \colhead{($c$)} & \colhead{($M_\odot$)} & \colhead{($M_\odot$)} & \colhead{} & \colhead{} & \colhead{} \\
}
\startdata
SN 2011kl & 0.68 & $1.9_{-0.41}^{+0.73}$ & $79_{-22}^{+14}$ & $6900_{-590}^{+630}$ & $0.07_{-0.03}^{+0.06}$ & $0.19_{-0.04}^{+0.07}$ & $11_{-0.86}^{+0.91}$ & $0.05_{-0}^{+0}$ & $1.60_{-0.03}^{+0.03}$ & $1.10_{-0.03}^{+0.03}$ & $19.00_{-0.02}^{+0.02}$ \\
CSS161010 & 0.03 & $0.18_{-0.04}^{+0.09}$ & $17_{-4.7}^{+10}$ & $9400_{-2600}^{+1600}$ & $0.84_{-0.22}^{+0.11}$ & $0.45_{-0.09}^{+0.04}$ & $13_{-0.37}^{+0.79}$ & $0.06_{-0}^{+0}$ & \nodata & \nodata & \nodata \\
AT2018cow & 0.01 & $0.52_{-0.11}^{+0.17}$ & $25_{-5.5}^{+6.3}$ & $9900_{-190}^{+81}$ & $0.59_{-0.05}^{+0.06}$ & $0.5_{-0.01}^{+0}$ & $14_{-0.13}^{+0.16}$ & $0.06_{-0}^{+0}$ & \nodata & \nodata & \nodata \\
AT2018lug & 0.27 & $5.8_{-3.9}^{+5.6}$ & $46_{-26}^{+34}$ & $3100_{-2000}^{+3100}$ & $0.31_{-0.2}^{+0.2}$ & $0.31_{-0.27}^{+0.14}$ & $24_{-3.5}^{+15}$ & $0.09_{-0.01}^{+0.03}$ & \nodata & \nodata & \nodata \\
AT2020mrf & 0.14 & $3.1_{-1.3}^{+2.5}$ & $63_{-23}^{+20}$ & $4200_{-1200}^{+1300}$ & $0.67_{-0.26}^{+0.23}$ & $0.14_{-0.04}^{+0.08}$ & $15_{-2.2}^{+3.7}$ & $0.07_{-0.01}^{+0.01}$ & \nodata & \nodata & \nodata \\
AT2020xnd\tablenotemark{b} & 0.24 & $0.58_{-0.33}^{+3.2}$ & $36_{-20}^{+36}$ & $1800_{-1300}^{+3300}$ & $0.27_{-0.14}^{+0.11}$ & $0.41_{-0.17}^{+0.06}$ & $18_{-0.99}^{+1.1}$ & $0.07_{-0}^{+0}$ & \nodata & \nodata & \nodata \\
MUSSES2020J & 1.1 & $7.4_{-3.2}^{+3.6}$ & $59_{-38}^{+28}$ & $8600_{-1600}^{+1000}$ & $0.84_{-0.22}^{+0.12}$ & $0.32_{-0.07}^{+0.09}$ & $28_{-4.8}^{+5.3}$ & $0.1_{-0.01}^{+0.01}$ & \nodata & \nodata & \nodata \\
AT2022tsd & 0.26 & $4.3_{-2.1}^{+5}$ & $61_{-28}^{+26}$ & $7300_{-3300}^{+2000}$ & $0.54_{-0.37}^{+0.32}$ & $0.23_{-0.13}^{+0.17}$ & $24_{-8.2}^{+9.1}$ & $0.09_{-0.02}^{+0.02}$ & \nodata & \nodata & \nodata \\
AT2023fhn & 0.24 & $1.6_{-1.1}^{+5}$ & $51_{-29}^{+30}$ & $4100_{-2600}^{+3600}$ & $0.24_{-0.17}^{+0.3}$ & $0.26_{-0.15}^{+0.15}$ & $26_{-4.6}^{+5.9}$ & $0.09_{-0.01}^{+0.01}$ & \nodata & \nodata & \nodata \\
AT2023hkw & 0.34 & $0.74_{-0.49}^{+3.1}$ & $43_{-23}^{+34}$ & $2600_{-1800}^{+3500}$ & $0.32_{-0.2}^{+0.22}$ & $0.33_{-0.11}^{+0.11}$ & $17_{-0.92}^{+2.3}$ & $0.07_{-0}^{+0.01}$ & \nodata & \nodata & \nodata \\
AT2023vth & 0.07 & $1.4_{-0.88}^{+3.7}$ & $52_{-29}^{+32}$ & $2900_{-2000}^{+3400}$ & $0.38_{-0.13}^{+0.23}$ & $0.34_{-0.06}^{+0.07}$ & $34_{-4.3}^{+3}$ & $0.11_{-0.01}^{+0.01}$ & \nodata & \nodata & \nodata \\
AT2024aehp & 0.17 & $7.6_{-0.85}^{+0.7}$ & $88_{-7.3}^{+7.3}$ & $3500_{-160}^{+180}$ & $0.03_{-0.01}^{+0.05}$ & $0.44_{-0.08}^{+0.04}$ & $17_{-0.95}^{+0.83}$ & $0.07_{-0}^{+0}$ & \nodata & \nodata & \nodata \\
AT2024qfm\tablenotemark{b} & 0.23 & $0.51_{-0.27}^{+1.9}$ & $41_{-23}^{+36}$ & $1900_{-1400}^{+4600}$ & $0.56_{-0.27}^{+0.21}$ & $0.42_{-0.13}^{+0.06}$ & $18_{-1.4}^{+1.8}$ & $0.07_{-0}^{+0}$ & \nodata & \nodata & \nodata \\
AT2024wpp\tablenotemark{b} & 0.09 & $1.3_{-0.74}^{+3.9}$ & $49_{-30}^{+36}$ & $530_{-300}^{+760}$ & $0.12_{-0.02}^{+0.03}$ & $0.23_{-0.01}^{+0.01}$ & $37_{-0.45}^{+0.55}$ & $0.12_{-0}^{+0}$ & \nodata & \nodata & \nodata \\
\enddata
\tablenotetext{a}{Derived.}
\tablenotetext{b}{Light curve is dominated by the ``fast" component though peak inferences on ``slow" component are likely unreliable.}
\end{deluxetable}
\end{longrotatetable}
\clearpage

\begin{deluxetable*}{lcccccccccc}
\tablecaption{ULGRB Host Galaxy Stellar Population Properties \label{tab:host}}
\tablehead{
\colhead{Object} &
\colhead{RA (J2000)} &
\colhead{Dec (J2000)} &
\colhead{$z$} &
\colhead{$\log(M_*/M_\odot)$} & 
\colhead{$t_m$ [Gyr]} &        
\colhead{SFR [$M_\odot$~yr$^{-1}$]} &
\colhead{$\log({\rm sSFR})$ [yr$^{-1}$]} &
\colhead{$\log(Z_*/Z_\odot)$} &  
\colhead{$A_V$ [mag]}
}
\startdata
GRB\,101225A & 00:00:47.68 & +44:36:01.6 & 0.847 & $8.33^{+0.51}_{-0.48}$ & $\cdots$ & $\cdots$ & $\cdots$ & $\cdots$ & $\cdots$  \\
GRB\,111209A & 00:57:22.66 & $-$46:48:03.6 & 0.677 & $8.81^{+0.33}_{-0.37}$ & $1.19^{+1.2}_{-0.87}$ & $0.6^{+0.95}_{-0.31}$ & $-9.07^{+0.59}_{-0.34}$ & $-0.47^{+0.48}_{-0.48}$ & $1.06^{+1.02}_{-0.65}$  \\
GRB\,121027A & 04:14:23.450 & -58:49:47.17 & 1.773 & $9.93^{+0.44}_{-0.45}$ & $0.5^{+0.56}_{-0.38}$ & $18.37^{+33.48}_{-7.78}$ & $-8.66^{+0.62}_{-0.33}$ & $-0.36^{+0.42}_{-0.42}$ & $0.63^{+0.99}_{-0.37}$  \\
GRB\,130925A & 02:44:42.38 & -26:09:15.8 & 0.348 & $9.13^{+0.32}_{-0.24}$ & $0.27^{+1.65}_{-0.2}$ & $4.74^{+9.57}_{-3.2}$ & $-8.38^{+0.57}_{-0.84}$ & $-0.39^{+0.51}_{-0.74}$ & $1.92^{+1.33}_{-0.8}$\\
GRB\,250702B & 18:58:45.57 & -07:52:26.2 & 1.036 & $10.72^{+0.22}_{-0.28}$ & $1.78^{+2.01}_{-1.14}$ & $14.14^{+34.81}_{-14.14}$ & $-9.57^{+0.75}_{-4.65}$ & $0.07^{+0.16}_{-0.32}$ & $2.82^{+1.26}_{-1.41}$ \\
\enddata
\tablecomments{
The redshifts, positions, stellar masses ($M_*$), mass-weighted ages ($t_m$), present-day SFRs and sSFRs, stellar metallicities ($Z_*$), and dust extinction ($A_V$) for the five ULGRB host galaxies in our sample. Given that the host of GRB~101225A was only detected in two photometric filters, we can only robustly estimate a stellar mass for its host.
}
\end{deluxetable*}

\begin{deluxetable}{lccccc}{}
\tabletypesize{\scriptsize}
\tablewidth{0pt}
\tablecaption{SNLC shock-heating model parameters for the SN~2011kl bolometric comparison. \label{tab:snlc_models}}
\tablehead{
\colhead{Model} & \colhead{$f_{\rm SBOE}$} & \colhead{$f_{\rm SBOM}$} & \colhead{$m_{\rm shell}$} & \colhead{$r_{\rm shell}^{\rm in}$} & \colhead{$r_{\rm shell}^{\rm out}$} \\
\colhead{} & \colhead{} & \colhead{} & \colhead{($M_\odot$)} & \colhead{(cm)} & \colhead{(cm)} 
}
\startdata
a & 5\% & 1\% & 1.5 & $2\times10^{15}$ & $7.5\times10^{15}$ \\
b & 1\% & 1.5\% & 1.5 & $10^{15}$ & $4\times10^{15}$ \\
c & 1\% & 2\% & 1.5 & $5\times10^{14}$ & $4\times10^{15}$ \\
d & 1\% & 2\% & 1.5 & $5\times10^{14}$ & $2.5\times10^{15}$ \\
e & 1\% & 2\% & 4.0 & $5\times10^{14}$ & $2.5\times10^{15}$ \\
f & 1\% & 2\% & 2.0 & $5\times10^{14}$ & $2.5\times10^{15}$ \\
g & 1\% & 1.5\% & 2.0 & $5\times10^{14}$ & $2.5\times10^{15}$ \\
\enddata
\end{deluxetable}

\software{\texttt{astropy} \citep{astropy:2013,astropy:2018,astropy:2022,astropy_17756022}, \texttt{matplotlib} \citep{Hunter:2007}, \texttt{numpy} \citep{numpy}, \texttt{python} \citep{python}, \texttt{scipy} \citep{2020SciPy-NMeth,scipy_20615351}, \texttt{corner.py} \citep{corner-Foreman-Mackey-2016,corner.py_14209694}, \texttt{Cython} \citep{cython:2011}, \texttt{emcee} \citep{emcee-Foreman-Mackey-2013,emcee_10996751}, \texttt{h5py} \citep{collette_python_hdf5_2014,h5py_7560547}, \texttt{FrankenBlast} \citep{frankenblast}, \texttt{Prospector} \citep{jlc+2021}; \texttt{FSPS} and \texttt{python-FSPS} \citep{FSPS_2009, FSPS_2010}
}


\bibliographystyle{aasjournal}
\bibliography{bibliography}

\end{document}